\documentstyle[eqsecnum,aps]{revtex}

\def\aut#1{#1}

\def\ins#1{}

\def\iems#1{}
\def\comment#1{}

\def\aut#1{#1}

\def\cm#1{}

 \def\lfrac#1#2{{{{#1}/{#2}}}}

\begin{document}
\setcounter{figure}{0}
\Roman{figure}

\title{{
Experimental Test for Absence of $R$-Term\\ in Schr\"odinger
Equation in Curved Space
}}
\author{
 H. Kleinert%
 \thanks{Email: kleinert@physik.fu-berlin.de\,, 
URL:
http://www.physik.fu-berlin.de/\~{}kleinert \hfill
} }
\address{Institut f\"ur Theoretische Physik,\\
Freie Universit\"at Berlin, Arnimallee 14,
14195 Berlin, Germany}
\maketitle
\begin{abstract}
We point out that
the presence of a term proportional to
the scalar curvature in the Schr\"odinger equation
in curved space can easily be detected
in atomic spectra with Russel-Saunders coupling
by a violation of the
Land\'e interval rule for adjacent levels
$ E_{J}-E_{J-1}\propto J$.
\end{abstract}
\maketitle

\section{Introduction}

In 1957, De Witt \cite{DW} presented a derivation
 of the Schr\"odinger
equation  of a point particle in curved space from Feynman's
path integral which contained an extra term proportional
to the scalar curvature $R$ in the Hamilton operator:
~\\[-8mm]
\begin{equation}
  \left( - \frac{\hbar^2  \Delta ^2}{2M}  + c\, \hbar^2 \,R\right)
    \psi (q,t) = i\hbar \partial_t \psi (q,t).
\label{1}\end{equation}
~\\[-2mm]
As usual, $M$ is the particle mass and
$ \Delta$
is
the Laplace-Beltrami operator.
The constant $c$ depends on the way the equation
is derived. This was initially
nonunique, allowing apparently
for different
ways of time slicing of the action
which
led to different values:
$c = \lfrac{1}{6}, c= \lfrac{1}{12}, \dots , c= \lfrac{1}{8}$.
The subject is discussed in detail in  textbook \cite{PI},
where
the ambiguity was
resolved with the help of a simple novel
equivalence principle requiring the
absence of the extra $R$ term [i.e. $c=0$ in (\ref{1})].
The new principle is deduced from the fact that spaces
with curvature (and torsion) can be reached by
a nonholonomic, and thus
multivalued coordinate transformation,
from a flat space \cite{GRG}. This proceeds by complete
analogy with the time-honored Volterra description
of defects in crystals \cite{GFCM} and
  of vortices in superfluids \cite{Cam}. It is also
analogous to Dirac's
generation of string-like magnetic flux tubes by performing
multivalued gauge transformation on the electromagnetic
vector potential  \cite{GRG}.

There seems to exist a wide-spread belief that
it is presently impossible to verify experimentally
the absence of an extra $R$-term. This is caused by the fact
that one usually thinks of $R$ as the curvature of
spacetime in general relativity, which is too small to be
detected in a quantum mechanical systems, since this
extends over
a too small spatial region.
It is the purpose of this note to point out that
there exists a completely different
possibility for experimental
verification in atomic physics.
Atoms whose shell of valence electrons is nearly
filled or almost empty, the spins and angular
momenta are coupled according to the Russell-Saunders scheme
and have an energy described by a Hamilton operator
~\\[-8mm]
\begin{equation}
\hat H = \frac{1}{2I_L} \hat{\bf L}^2 + \frac{1}{2I_S} \hat{\bf S}^2
 + \frac{1}{I_{LS}} \hat{\bf L} \cdot \hat{\bf S},
\label{HO}\end{equation}
~\\[-2mm]
where $\hat{\bf L} $ and $ \hat{\bf S}$ are the operators of total angular
momentum and spin of the electrons, $\hat{\bf L} \cdot \hat{\bf S}$
is the spin-orbit interaction,
and $I_L,\,I_S$ are moments of inertia.
The interaction constant $I_{LS}$ is positive for low and
negative for high filling \cite{LL}.
By calculating the expectation values $\hat{\bf L} \cdot \hat{\bf S}$
in states of total angular momentum $J$ via
the commutation rules of the rotation group one finds
the eigenvalues
\begin{equation}
  2\,\hat{\bf L} \cdot \hat{\bf S} =
  \hat{\bf J}^2 -
  \hat{\bf L}^2-
  \hat{\bf S}^2
 =
 J(J+1) - L(L+1) - S(S+1),
\label{@}\end{equation}
~\\[-6mm]
which lead to the famous
 {\em Land\'e
 interval rule\/} for adjacent levels
 \cite{LL}:
\begin{equation}
  E_J-
   E_{J-1} = \frac{1}{I_{LS}}  J.
\label{@Lan}\end{equation}
~\\[-8mm]

~\\[-1.cm]
\section{Curved-Space Formulation of Hamilton Operator}
We now show that this well-obeyed rule would be
destroyed by an additional $R$-term in the Schr\"odinger
equation (\ref{1}) with any of the theoretically proposed
constants $c\neq 0$. The argument goes as follows. Since
Schr"dinger quantum
mechanics is independent of the coordinates used to describe a
system,
we may consider, for even total spin $S$, the Hamilton
operator (\ref{HO}) as the quantized version of a classical
system in which two point particles
of masses $I_L$ and $I_S$ move on a unit sphere. Their
Lagangian
 has the general form
\begin{equation}
L = \frac{1}{2}\, g^{\mu \nu } \dot{q}^\mu  \dot{q}^ \nu,
\label{LA}\end{equation}
~\\[-5mm]
where $(q^1, q^2) = ( \theta, \varphi)$ and
 $(q^3, q^4) = ( \theta', \varphi')$ are spherical angles,
and $g_{\mu \nu }$ is the
dynamical metric.
%
%
The associated Hamiltonian is
~\\[-8mm]
\begin{equation}
H = \frac{1}{2}\, g^{\mu \nu } {p}_\mu  {p}_\nu,
\label{gpp}\end{equation}
where
$p_1=p_\theta=\dot \theta,~p_2=p_\phi=\sin ^2\theta\,\dot \varphi$
and $p_3=p_{\theta'}=\dot \theta',~p_4=p_{\phi'}=\sin ^2\theta'\,\dot \varphi'$
are the canonical momenta
and $g^{\mu \nu }$ is the inverse
of the metric $g_{\mu \nu }$.
\begin{equation}
g^{\mu \nu }= \frac{1}{I}
\left(
\begin{array}{lc}
1&0\\0&\sin^{-2}\theta
\end{array}
\right)
\label{@}\end{equation}
To find the metric associated with the
Hamiltonian operator (\ref{HO}), we consider
for a moment
a single point particle of mass $I$ moving on a unit sphere with the
Cartesian coordinates ${\bf x} = (\sin \theta \cos \varphi, \sin
\theta \sin \varphi, \cos \theta)$.
Its angular momentum has the components
\begin{eqnarray}
 L_1 & = &
- \sin \varphi \,\dot \theta - \sin \theta   \cos \theta
  \cos \varphi \,\dot\varphi
=- \sin \varphi \,p_\theta -   \cot \theta
  \cos \varphi \,p_\varphi
, \nonumber\\
 L_2 & = &~~\,
\cos \varphi \,\dot{\theta} - \sin \theta
  \cos \theta \sin \varphi \,\dot \varphi
=~~\,
\cos \varphi \,p_{\theta} -
  \cot \theta \sin \varphi \,p_\varphi
, \nonumber\\
 L_3 & = &~~\,
 \sin^2 \theta \,\dot\varphi~~~~~~~~~~~~~~~~~~~~~~~~\!\, = ~~\,
 p_\varphi .
\label{}\end{eqnarray}
The Hamiltonian can therefore be written as
\begin{equation}
    H =
\frac{1}{2I}\left(p_\theta^2+\frac{1}{\sin^2\theta}p_\varphi^2\right)
  =\frac{1}{2}\, {\bf L}^2,
\label{@}\end{equation}
which has the form (\ref{gpp}) with
the inverse  metric
\begin{equation}
g^{\mu \nu }= \frac{1}{I}
\left(
\begin{array}{lc}
1&0\\0&\sin^{-2}\theta
\end{array}
\right)
\label{@}\end{equation}
By parametrizing the even angular momentum ${\bf S}$ likewise
and forming a combination as in (\ref{HO}), we obtain
a combined Hamiltonian
of the form
(\ref{gpp}) in four dimensions
with the  following
inverse metric
%
\begin{eqnarray}
g^{\theta\theta} &=&\frac{1}{I_L}
 ,~~~
g_{\theta\phi} = 0,~~~
g_{\theta\theta'} =
{g_{LS}} \cos (\varphi-\varphi'),~~~
g_{\theta\phi'} =
  \frac2{I_{LS}}
\cot \theta'
\sin (\varphi - \varphi ')
,\nonumber \\
g_{\phi  \phi}&\!\!\!=\!\!\!&
 \frac{1}{I_L}\left(1+ \cot^2 \theta \right),~~~
g_{\phi  \theta'} =  - \frac2{I_{LS}}
\cot \theta
 \sin (\varphi - \varphi ')
,~~~
g_{\phi  \phi'} =\frac{2}{I_{LS}}
{\cot \theta \cot \theta '}
    \cos(\varphi - \varphi')
\nonumber \\
g_{\theta'\theta'} &=&
  \frac{1}{I_S}  ,~~~
g_{\theta'\phi'} =0,~~~
g_{\phi'\phi'} =\frac{1}{I_S}\left(1+ \cot^2 \theta' \right)
.
\end{eqnarray}
For this we now calculate the Laplace-Beltrami operator
$ \Delta  = g^{-1/2} (\partial_\mu g^{\mu \nu } g^{1/2} \partial_ \nu )$
where
$g$ is the determinant of $g_{\mu \nu }$,
 and find a Schr\"odinger equation
(\ref{1}).
This can be written in the form
\begin{equation}
\hat H\psi(q,t)=i\hbar \psi(q,t)
\label{@}\end{equation}
with a Hamilton operator of the form (\ref{HO}),
and differential operators for the angular momenta,
\begin{eqnarray}
\hat  L_1 & = & ~~\,i\hbar \left(\sin \varphi \,\partial _\theta +\cot\theta\cos\varphi\,\partial _\varphi\right),\nonumber \\
\hat  L_1 & = & -i\hbar \left(\cos \varphi \,\partial _\theta -\cot\theta\sin\varphi\,\partial _\varphi\right),\nonumber \\
 \hat L_3 & = &-i\hbar \,\partial _\varphi ,
\label{}\end{eqnarray}
and a similar representation for $\hat S_{1,2,3}$
in terms of $\theta'$ and $\varphi'$.
Thus the Hamilton operator (\ref{HO})
coincides with the Schr\"odinger equation (\ref{1})  in curved space,
{\em without\/} an extra $R$-term.
~\\[-.7cm]

\section{Effect of Extra $R$-Term on Atomic Levels}
~\\[-.7cm]
Let us now see how an extra $R$-term
would change the spectrum of $\hat H$.
The calculation of $R$
in this four-dimensional space is quite tedious,
but can easily be
done with the help of the algebraic computer program
{\tt Reduce} (using the package {\tt excalc}).
The final result is stated most compactly for
the
limiting  case where
 $I^{-1}_{LS}$ is much smaller than
$I^{-1}_L+I^{-1}_L$.
Then we find
~\\[-8mm]
\begin{eqnarray}
 R & = &2\left(
\frac{1}{I_L}+
\frac{1}{I_S}
\right)+
\frac{16}{3}\frac{{I_L}+{I_S}}{{I_{LS}}}
\left(f-1\right)
,
 \label{R2}
\end{eqnarray}
~\\[-.5cm]
with
~\\[-.5cm]
\begin{equation}
 f = 4-6 \sin^2 \beta   ,
\label{f}\end{equation}
~\\[-3mm]
where $ \beta $
is the relative angle between ${\bf x}$ and ${\bf x}'$.
The function $f$ is equal to $ \sqrt{5}/16 \pi$
times the Clebsch-Gordan combination of the
spherical harmonics
$Y_{1m}(\theta,\varphi)$
and $Y_{1m'}(\theta',\varphi')$
with a total angular momentum $(J,M)=(2,0)$.
This makes it
 straight-forward
to calculate the matrix elements of $f$ between
states of total angular momentum $J$.
The results for
$S = 1 $ and $L= 1 ,2, 3$
are listed in Table \ref{tb1}.
The table shows that the Land\'e interval rule would
be violated
if an extra $R$-term were present
in the Schr\"odinger equation (\ref{1}).
We expect that this violation would have been
noticed before in the analysis of atomic spectral lines
if $c$ had any of the theoretically
expected nonzero constants $c$.

Nevertheless it would be interesting to reexamine
the validity of the Land\'e interval
rule and derive from this
upper bounds for the constant $c$.

Note that in spite of the small factor $\hbar^2$ in front of $R$
in the
 Schr\"odinger
equation
(\ref{1}),
the shift of levels by (\ref{R2}) has the same order of magnitude
as the Land\'e intervals if $c$ and $a,\varepsilon$ are of order unity.
This is due to the factors $\hbar ^2$ in
$\hat{{\bf L}}^2$,
$\hat{{\bf S}}^2$,
and
$\hat{{\bf L}}\cdot
\hat{{\bf S}}$.
~\\[-.7cm]
\section{Acknowledgment}
~\\[-.7cm]
The author thanks
Drs.~A. Chervyakov and H.J.~Schmidt for discussions and
Dr.~Eberhard~Schruefer for helping him to make
to run his {\tt Reduce} program.

~\\[-1.3cm]

~\\[-1.3cm]
\begin{table}[tbhp]
\caption[]{Diagonal matrix elements
of function $f$ of
Eq.~(\ref{f}), determining the violation of the Land\'e interval rule
by the presence of an extra $R$-term in the Schr\"odinger equation
(\ref{1}) by additional level differences.
The change of the proportionality factor $1/I_{LS}$
of
$J$
in Eq.~(\ref{@Lan})
is listed in the last column.{} }
\begin{tabular}{ccc|r|r|}
$S$ & $L$ & $J $&$\langle J| f|J\rangle $&$(E_J-E_{J-1})/J$ \\
\hline
1 & 1 & 0 & 8/5&\\
1 & 1 & 1 & -4/5&-60/25\\
1 & 1 & 2 & 4/25&12/25\\
\hline
1 & 2 & 1 &  4/5&\\
1 & 2 & 2 & -4/5&-28/35\\
1 & 2 & 3 & 8/35&12/35\\
\hline
1 & 3 & 2 & 16/25& \\
 1 & 3 & 3 & -4/5 &-72/150\\
1 & 3 & 4 & 4/15&-20/150\\
\hline
2 & 2 & 0 & 8/7 &\\
2 & 2 & 1 & 4/7& -84/147\\
2 & 2 & 2 & -12/49&-60/147 \\
2 & 2 & 3 & -32/49& -20/147\\
2 & 2 & 4 & 16/49 & 36/147
\end{tabular}
\label{tb1}\end{table}

\end{document}